%% file: ver2_astroph.tex
\documentstyle[12pt,aasms4,epsf]{article}
\received{00  2001}
\accepted{00 2001}
\centerline{ Submitted to the Editor of ApJ Letter
 on July 16, 2001}

\def\ee #1 {\times 10^{#1}}          
\def\ut #1 #2 { \, \textrm{#1}^{#2}} 
\def\u #1 { \, \textrm{#1}}          

\def\kms    {\hbox{km{\hskip0.1em}s$^{-1}$}}    
\def\msol   {\hbox{$M_\odot$}}                  

\def\kms    {\hbox{km{\hskip0.1em}s$^{-1}$}}    
\def\etal   {{\it et al. }}                     

\begin{document}
\title{Detection of X-ray Emission from
the Arches Cluster Near the Galactic Center}

\author{F. Yusef-Zadeh}
\affil{Department of Physics and Astronomy, Northwestern University,
Evanston, Il. 60208 (zadeh@northwestern.edu)}

\author{C. Law}
\affil{Harvard-Smithsonian Center for Astrophysics, Cambridge, MA. 02138
 (claw@head-cfa.harvard.edu)}

\author{M. Wardle}
\affil{School of Physics, University of Sydney, NSW 2006, Australia
(wardle@physics.usyd.edu.au)}

\author{Q. D. Wang}
\affil{University of Massachusetts, Amherst, MA 01003-4517
 (wqd@hnr.astro.umass.edu)}

\author{A. Fruscione}
\affil{Harvard-Smithsonian Center for Astrophysics, Cambridge, MA. 02138
 (antonell@head-cfa.harvard.edu)}

\author{C.C. Lang}
\affil{University of Massachusetts, Amherst, MA 01003-4517
 (clang@ocotillo.astro.umass.edu)}

\author{A. Cotera}
\affil{University of Arizona, Steward Observatory, Tucson, AZ 85721
 (cotera@as.arizona.edu)}

\begin{abstract}

The Arches cluster is an extraordinarily compact massive star cluster
with a core radius of about 10$''$ ($\sim$0.4 pc) and consisting of
more than 150 O star candidates with initial stellar masses greater
than 20~M$_\odot$ near G0.12-0.02.  X-ray observations of the radio
Arc near the Galactic center at l$\sim0.2^0$ which contains the Arches
cluster have been carried out with the Advanced CCD Imaging
Spectrometer (ACIS) on board Chandra X-ray Observatory. 
We report the detection of two
X-ray sources from the Arches cluster embedded within a bath of
diffuse X-ray emission extending beyond the edge of the cluster to at
least 90$''\times60''$ (3.6 pc $\times$ 2.4 pc).  The brightest
component of the X-ray emission coincides with the core of the cluster
and can be fitted with two-temperature thermal spectrum with a soft
and hard component of 0.8 and 6.4 keV, respectively.  The core of
the
cluster coincides with several ionized stellar wind sources that have
previously been detected at radio wavelengths, suggesting that the
X-ray emission from the core arises from stellar wind sources.  The
diffuse emission beyond the boundary of the cluster is discussed in
the context of combined shocked stellar winds escaping from the
cluster. We argue that the expelled gas from young clusters such 
as the Arches cluster may
be responsible for the 
hot and extended X-ray emitting gas detected throughout 
the inner degree of  the  Galactic center. 

\end{abstract}


\vfill\eject

\section{Introduction}
Near-IR observations of the Galactic center region have recently identified
two clusters of young and massive stars embedded within the radio Arc at
galactic longitude l$\sim0.2^0$ lying within 25 pc in projection from the
Galactic center at a distance of 8.5 kpc. (Nagata et al. 1995; Cotera et
al. 1996; Serabyn, Shupe, \& Figer 1998; Figer \etal 1999). The two stellar
clusters, which are responsible for ionizing the thermal components of the
radio Arc, show similar characteristics to the IRS 16 cluster at the
Galactic center (Simons, Hodapp \& Becklin 1990; Allen \etal 1990, Krabbe
\etal 1991).

Zwart \etal (2001) have recently predicted that the Galactic center region
may contain a large of number of young and massive clusters similar to the
Arches, Quintuplet and IRS 16 clusters. Understanding their nature is an
important step towards determining the rate of massive star formation in
this unique region of the Galaxy.

The Arches cluster has an angular
size of $\sim15''$ with a peak density of $3\times10^5$ \msol
pc$^{-3}$ in the inner 9$''$ (0.36 pc) (Figer \etal 1999).
This cluster is one of the densest known young clusters in the local group
of 
galaxies  with densities similar to 
R136, the central cluster of 30 Dor in the Large Magellanic Cloud and  
NGC 3603;  the Trapezium cluster in Orion has  a
density of 7$\times10^4$  \msol pc$^{-3}$ (Brandl \etal 1996; Wang 1999; 
McCaughrean \&
Stauffer 1994; Brandl \etal  1999). The
Arches cluster has an estimated age of 1--3 Myr and shows a flat mass
function as compared to other young clusters
(Figer \etal 1999).  The Quintuplet cluster
adjacent to the Sickle is less compact, and is 3--5 Myr old
(Figer, McLean and Morris 1999).
Radio continuum emission from individual stars
in both clusters has recently been detected (Lang et al.  1999; 2001).  
Radio spectral index 
and near-IR spectral type of several stars in the Arches
cluster are consistent with ionized stellar winds arising from
mass-losing WN and/or Of stars with mass-loss rates $\approx
(1-20)\times10^{-5}$ \msol\ yr$^{-1}$ and lower limits to their winds' 
terminal
velocities range  between 800 and 1200 \kms (Cotera \etal 1996).

\section{Observations}

The X-ray imaging of the inner degree of the Galactic center region
dates back to Einstein and SL2-XRT, ROSAT, ASCA observations where a
number of point sources and diffuse emission were detected (Watson
\etal 1981; Skinner \etal 1987; Predehl and Tr\"umper 1994; Koyama
\etal 1996). 
The Advanced CCD Imaging Spectrometer (ACIS) on board the Chandra X-Ray
Observatory (Weisskopf \etal 1996; Garmire \etal 2000) was used to
observe the radio Arc on July 7, 2000 for a total observing time 51
kseconds.  The observation was made in the ACIS-I configuration, with
a nominal aim-point toward the linear filaments of the radio Arc (epoch
2000)
$\alpha=17^h 46^m 22^s, \delta=-28^0 51'36.4''$.  In total, six of the
ACIS CCDs were readout:  the imaging array, I0-I3, and two of the
spectroscopy array, S2 and S3.  Only the data from the imaging array,
particularly the I1 chip, will be presented here.
The data were processed using the Chandra X-ray Center's
CIAO software package (Noble \etal 2000).
The images are 
effectively flat-fielded:  instrument response and effective area are
divided out in the form of an ``exposure map''.


\section{Results}

Figure 1a shows a flux image of X-ray emission detected in ACIS's I$_0$ to
I$_3$ detectors with prominent sources labeled. Two sources located to the
northwest of the image coincide with the Arches cluster. These   
sources  avoid  the
strong diffuse X-ray emission arising from the Galactic plane  of 
the Galactic center region.
We note a
prominent diffuse structure on a scale of about 2$'\times2'$ centered near
(epoch 2000) $\alpha=17^h 46^m 18^s, \delta=-28^0 54'$. This 
diffuse feature shows 
two linear X-ray structures.  
Using the Very
Large Array of the National Radio Astronomy Observatory\footnote{The
National Radio Astronomy Observatory is a facility of the National Science
Foundation, operated under a cooperative agreement by Associated
Universities, Inc.} a $\lambda$20cm continuum image with a resolution of 
10.7$''\times10''$ (Yusef-Zadeh, Morris and Chance 1984) 
is compared with
identical region to that of Figure 1a and is displayed in Figure 1b where
Sgr A complex is located to the SW corner.  The northern X-ray filament
appears to be situated at the inner boundary of a number of nonthermal
radio filaments running perpendicular to the Galactic plane. 

Figure 2a shows contours of adaptively smoothed X-ray emission in a smaller
region around the Arches cluster, with five components being
identified and labeled as A1--A5.  The crosses show the
positions of the  compact radio sources which are due to free-free
emission from ionized stellar winds (Lang \etal 2001).  The brightest
component of X-ray emission (A1) is centered on the SW corner of the
densest part of the core of the Arches cluster where the largest
concentration of ionized stellar wind radio sources has been detected.
The near-IR counterparts to radio sources are identified as either WN
or Of spectral types (Nagata \etal 1995; Cotera \etal 1996). 

A close-up view is displayed in Figure 2b where the X-ray contours
have been superimposed on a grayscale image of the stellar sources
obtained with NICMOS on the HST (Figer \etal 1999). 
The brightest \emph{radio} source, an Of/WN9 
star (star no. 8 in Nagata \etal 1995 and Cotera \etal 1996) 
estimated to have a mass-loss rate of 1.7$\times10^{-4}$ \msol
yr$^{-1}$ (Lang \etal 2001),  lies very close to the geometric
center of A1.  These facts strongly suggest that A1 is associated with
the
core of the Arches cluster.  The second X-ray component (A2) lies
about 10$''$ NW of A1 and appears to be located outside the core of
the cluster but still close to the outer boundary.  No radio emitting
stellar sources have been detected toward A2.  
The peak of A2 coincides within 1$''$ with a star classified 
an emission line star, WN7, WN8 or Of4 (Cotera \etal 1996, Blum \etal 
2001).
The largest scale
component of X-ray emission in Figure 2a is a diffuse ovoid
feature (A3)
which
envelops both A1 and A2, and is elongated at a position angle 
of 102$^0$ E of N from the Galactic plane.  With dimensions of
approximately 90$''\times 60''$ (3.6 $\times$ 2.4 pc), A3 extends well
beyond the edge of the cluster, which is $\la$ 15$''$ in diameter.
Two point-like (A4 and A5) lie  to
the east and west of the cluster at (J2000) $\alpha=17^h 45^m 55.84^s,
\delta=-28^0 49' 15.1''$ and (J2000) $\alpha=17^h 45^m 45.25^s,
\delta=-28^0 49' 26.1''$.  The nature of A4 and A5 as well as a
number of extended  and sharp features observed 
in high-resolution X-ray images  will be
discussed elsewhere.

Figures 3a,b and c show the observed and modeled spectra toward the A1, A2
and A3 components.  The spectra of A1 and A2 were fitted with a MEKAL
model (Mewe \etal 1985)  of a two-temperature thermal plasma.  The two
sources were
permitted to have different temperatures and volume emission measures,
but the abundances and absorbing column densities were constrained to
be identical.  Table 1 shows the parameters of the A1 and A2 model
fits and one $\sigma$ uncertainty 
and the associated 68\% confidence level. 
The spectra are best
fitted by thermal bremsstrahlung with two
temperatures at 0.8 and 6.4 keV for A1 and 0.9 and 5.8 keV for A2.
The respective X-ray luminosities (L$_X$) of A1 and A2 integrated
between 0.2 and 10 keV, give 3.3 and 0.8$\times10^{35}$ ergs s$^{-1}$
with a total L$_X\sim4.1\times10^{35}$ ergs s$^{-1}$.

The heavy element abundances in the X-ray emitting plasma range from 
1.6 to 3.8 solar, similar to other determinations in the Galactic 
center region (Maeda \etal 2001; Baganoff \etal 2001).  The absorption 
column of 12.4$^{+2.9}_{-2}\times10^{22}$ H cm$^{-2}$ toward the 
Arches cluster corresponds to a visual extinction of 
A$_V\sim69^{+16}_{-11}$ mag, similar to that inferred from Chandra and 
ROSAT observations toward Sgr A East and Sgr A$^*$ at the Galactic 
center (Maeda \etal 2001; Baganoff \etal 2001; Predehl and Tr\"umper 
1994).  There seems to be a systematic discrepancy between X-ray 
estimates of A$_V$ and the value $\sim 30$ inferred from near-IR 
observations toward the Arches cluster and the Galactic center 
(Serabyn \etal.  1998; Rieke and Lebofski 1985).  Note that the fits 
to the spectra are poorer (reduced chi-square $\sim 1.5$) when N$_H$ 
is set to $6\times 10^{22}$ H cm$^{-2}$, corresponding to A$_V=$ 30 
mag.

After the A1 and A2 components are subtracted the spectrum of A3 is 
fit by a single thermal bremsstrahlung of temperature 5.7$_{4.4}$ keV, 
an absorbing column of 10.1$^{13.5}_{7.9}\times10^{22}$ H cm$^{-2}$ 
and an additional Gaussian contributed by fluorescent Fe K/$\alpha$ 
6.4 keV line emission.  The upper bound to the temperature is 
unconstrained due to the low number of counts at high energies and the 
restricted energy band, although the estimated temperatures of A1, A2 
and A3 are the same within errors.  The total X-ray luminosity of A3 
between 0.5 and 10 keV excluding the 6.4 keV line is L$_X\approx 
1.6\times10^{34}$ ergs s$^{-1}$.

\section{Discussion}

The Arches cluster was positioned on the I1 chip, about $7'$ away from 
the telescope's aimpoint.  The shape and the size of the PSF at this 
location is fitted by an elliptical Gaussian with 
FWHM=4.4$''\times2.2''$ and PA=83$^0$.  Sources A1 and A2 are fitted 
by elliptical Gaussians with FWHM 8.0$''\times7.1''$ (PA=96$^0$) and 
7.9$''\times5.1''$ (PA=90$^0$), respectively, indicating that they are 
partially resolved.  Further observations are needed to confirm the 
true extent of A1 and A2.

The resulting distortion of A1 and A2 caused by the point spread 
function (PSF) means that the present observations cannot distinguish 
between compact emission associated with the collisions between the 
ionized winds from early-type stars paired in binary systems and more 
extended emission arising as a result of the interaction between winds 
from individual stars within the cluster.  Figure 2b suggests an 
association of the A1 component with the known stellar wind sources in 
the Arches cluster, and of A2 with the bright stellar source located 
at (J2000) $\alpha=17^h 45^m 50.27^s, \delta=-28^0 49' 11.64''$ which 
is identified as either a WN8 or an Of4 star (star number 1 in Cotera 
\etal 1996).  The X-ray emission from the collision of winds between 
early-type stars and early-type companions can account for the 
luminosity and the temperature of A1 and A2 (Stevens \etal 1992).  The 
X-ray emission from the cluster R136 is believed to be powered by a 
dozen colliding wind X-ray binaries.  The brightest binary system (CX5 
in Zwart \etal 2001) has X-ray luminosity of 2$\times10^{35}$ ergs 
s$^{-1}$ and a fitted temperature of 2.3 keV (Zwart, Walter and Lewin 
2001).

The A3 component is unlikely to arise from individual stellar sources 
or binary systems associated with the cluster as it extends far beyond 
the cluster boundary.  A more likely source is the shock-heated gas 
created by the collision of individual $\sim 1000$ km/s stellar winds 
in the dense cluster environment (Ozernoy, Genzel and Usov 1997; 
Cant\'o \etal 2000).  This gas is far too hot to be bound to the 
cluster, escaping as a supersonic wind provided the external pressure 
of the medium is not too high.  The electron density and mass of A3 
implied by our observations are 1.9 cm$^{-3}$ and 0.4 \msol, 
respectively.  Using the total mass loss rate $\sim 4\times10^{-4}$ 
\msol\ yr$^{-1}$ estimated from the radio continuum sources detected 
in the cluster (Lang \etal 2001), we estimate that the residence time 
of the gas in the A3 component is about 10$^3$ years.  As predicted by 
the model (Cant\'o et al.  2000), the required flow velocity of the 
wind, about 1200 \kms, is comparable to the stellar wind speeds 
determined from the spectra of cluster members (Cotera \etal 1996).  
The elongation of the X-ray emission from A3 perpendicular to the 
Galactic plane may be an indication that the cluster wind flow is 
confined more by the ISM in the galactic plane than normal to it.

The X-ray emission from the wind is dominated by the portion within 
the cluster boundary, thus either or both of the A1 and A2 components 
may be produced by the wind.  The total X-ray flux between 0.2 and 10 
keV from A1, A2 and A3 is $\sim5\times10^{35}$ erg s$^{-1}$ which is 
close to the X-ray luminosity of 6$\times10^{35}$ ergs s$^{-1}$, 
predicted by Cant\'o \etal (2000).  Their temperature estimate also 
agrees with the high temperature component of A1 (if a mean molecular 
weight of 2m$_H$ is assumed instead of 2/3 m$_H$).

Where does the hot gas go after it escapes from the cluster?  
Theoretical studies predict that at any one time the inner 200 pc of 
the Galaxy may harbor 50 clusters like the Arches with lifetimes of 
roughly 70 Myr (Zwart \etal 2001). 
 It is possible that the gas 
expelled from the Arches clusters and others like it contributes to 
the extremely hot diffuse gas found in the inner degree of the 
Galactic center.  The electron temperature of the hot gas is about 10 
keV, with an electron density of 0.3--0.4 cm$^{-3}$ and a total mass 
of about 2--4$\times 10^3$ \msol\ (e.g.  Yamauchi \etal 1990; Koyama 
\etal 1996).  The hot plasma at this temperature can not be confined 
by the gravitational potential in the Galactic center region, and if 
unhindered escapes the region on a time scale of $10^5$ years.  The 
required replenishment rate is therefore $\sim 3 \times 10^{-2}$ 
\msol\ yr$^{-1}$, which could be supplied by the gas lost from 50 
clusters like the Arches cluster in the inner 200 pc.  However, if the 
strength of the magnetic field in the inner degree is about one 
milliGauss, as a number of studies indicate (e.g.  Yusef-Zadeh.  
Morris and Chance 1984; Yusef-Zadeh and Morris 1987; Lang, Morris and 
Echevarria 1999), then the magnetic pressure is strong enough to 
confine much of the hot plasma (Koyama \etal 1996; Yusef-Zadeh \etal 
1997).  Under this assumption, the gas must be replenished on the 
cooling time of $\sim 10^8$ years, and the required input rate is 1000 
times less.  The confined gas can be accounted for if at any given 
time there is a single cluster similar to the Arches cluster within 
the inner degree of the Galactic center.

Searches for extended X-ray emission may prove 
useful in identifying young clusters near the Galactic 
Center.  It would be worthwhile to look for X-ray emission from 
young cluster candidates (e.g. Dutra and Bica 2000).

\input table1.tex

\begin{figure}
\caption{(1a) An  image  of Chandra's four ACIS-I chips
for the  0.5 to 8 keV band where diffuse emission
is detected throughout this region. 
We note 
 two linear X-ray structures
running perpendicular to
the  Galactic plane. 
A bright source  to  the north at the edge of the field
is identified with 1E1743.1--2843 (Watson \etal 1981)
whereas the  bright feature at the bottom right  corner
is associated with the Sgr A complex.
(1b) A  radio 
counterpart to an identical region of (a) but at $\lambda$20cm
with a spatial resolution of 10.7$''\times10''$. The nonthermal radio
filaments
are seen prominently in the direction perpendicular to the
galactic plane. 
The northern X-ray filament is situated at the inner boundary
of the nonthermal radio  filaments.}
\end{figure}

\begin{figure}
\caption{Left panel (2a) shows contours of adaptively smoothed X-ray
emission of five  components (A1--A5) of the Arches cluster. The crosses
show 
the positions of mass-losing stellar wind sources identified
in radio wavelengths. 
The total X-Ray flux of A4 and A5 are 1.4 and
0.18$\times10^{-12}$ ergs s$^{-1}$cm$^{-2}$, respectively.
Right panel (2b) shows contours of the
smoothed X-ray emission from A1 and A2
with stellar radio sources as crosses and are superimposed on the
near-IR stellar distribution of the Arches cluster.
}
\end{figure}

\vfill\eject

\begin{figure} 
\caption{The spectra and the fits to the A1, A2 and A3
components of the Arches cluster are shown in the top left (a) and top
right (b)  and bottom (c) panels, respectively.  The A1 and A2 fits model
with an absorbed, two-temperature, thermal plasma emission model whereas
the A3 fit model is an absorbed, one temperature, thermal plasma emission
model with a Gaussian at 6.4 keV corresponding to the fluorescent Fe
K/$\alpha$ emission.  The spectra of the A1 component of Arches cluster are
extracted from an elliptical region with an extent of 6$''$x11$''$.
  The
background subtracted  from the  is  taken
from the
region outside of the Arches cluster, in a roughly circular region 200$''$
across.  Sherpa, the fitting and modeling engine of CIAO, was used to fit
the A1 and A2 spectra.  We fit the spectra with ''xsvmekal'' (Mewe,
Groneschild and ven den Oord 1985;  a.k.a. ''vmekal'' in XSPEC), a thermal
bremsstrahlung plasma model with five emission lines of varying abundance
(Si, S, Ar, Ca, Fe).  The thermal model was also multiplied by an absorbing
column, ''xswabs''.  For the two-temperature fits to A1 and A2, the
elemental abundances and column densities of the first and second
components were linked together.  The thermal flux of absorbed
(absorption-corrected) spectrum when integrated between 0.2 and 10 keV give
4.8$\times10^{-13}$ (3.9$\times10^{-11}$) ergs cm$^{-2}$ s$^{-1}$ for A1
and 2.7$\times10^{-13}$ (9$\times10^{-12}$) ergs cm$^{-2}$ s$^{-1}$ for A2
covering a region of 5$''\times10''$ and 5.1$\times10^{-13}$
(1.8$\times10^{-12}$) ergs cm$^{-2}$ s$^{-1}$ for A3 covering a region of
74$''\times111''$ with A1 and A2 removed. 
}
\end{figure}

\end{document}

%% file: table1.tex
\input epsf
 
\def\k{km s$^{-1}$}
\def\pp{^{\prime\prime}}
\def\cm2{cm$^{-2}$}
\def\c3{cm$^{-3}$}


\begin{deluxetable}{lcrl}

\singlespace

\tablecaption{Best-fit Parameters to the Components of the
Arches Cluster\label{tbl3}}
\tablehead{
        \colhead{Source} &
        \colhead{Parameter} &
        \colhead{Best fit (error bars)}}
\startdata
A1 &kT(1)[keV] & 6.4 (-1.5, +2.8) &\nl
$''$ &normalization & 3.6$\times10^{-4}$ (-1.8$\times10^{-4}$,
+2.2$\times10^{-4}$) &\nl
$''$ &kT(2)[keV] & 0.8 (-0.2, +0.2) &\nl
$''$ &normalization & 8.3$\times10^{-3}$ (-5.5$\times10^{-3}$,
+2.3$\times10^{-2}$) &\nl
$''$ &chi-squared per degree of freedom & 0.98 &\nl

A2 &kT(1)[keV] & 1.0 (-0.4, +0.6) &\nl
$''$ &normalization & 2.0$\times10^{-3}$ (-1.3$\times10^{-3}$,
+1.0$\times10^{-2}$) &\nl
$''$ &kT(2)[keV] & 5.8 (-2.3, +4.7) &\nl
$''$ &normalization & 2.3$\times10^{-4}$ (-1.4$\times10^{-4}$,
+2.1$\times10^{-4}$) &\nl
$''$ &chi-squared per degree of freedom & 0.97 &\nl

A1, A2 &N$_{H} [10^{22}$ cm$^{-2}$]&12.4 (-2.0, +2.9)  &\nl
$ ''$ &Si/Si$\odot$ & 2.9 (-1.7, +3.7) &\nl
$''$ &S/S$\odot$   & 1.6 (-0.6, +1.0 ) &\nl
$''$ &Ar/Ar$\odot$  & 2.5 (-1.0, +1.3) &\nl
$''$ &Ca/Ca$\odot$  & 3.9 (-1.4, +1.7) &\nl
$''$ &Fe/Fe$\odot$  & 2.2 (-0.8, 1.7) &\nl

\enddata

\end{deluxetable}

